\pdfoutput=1
\documentclass[12pt]{nature}

\usepackage{graphicx}
\usepackage{graphics}
\usepackage{amssymb,amsmath}
\usepackage{wasysym}
\usepackage{times}
\usepackage{lineno}

\title{Imaging Viscous Flow of the Dirac Fluid in Graphene Using a Quantum Spin Magnetometer}

\author{Mark J.H. Ku$^{1,2,*}$, \,Tony X. Zhou$^{2,3,*}$,\, Qing Li$^{2}$, \,Young J. Shin$^{2,4}$,\,
Jing K. Shi$^{2}$, \,Claire Burch$^{5}$, \,Huiliang Zhang$^{1,2}$, \,Francesco Casola$^{1,2}$,\,
Takashi Taniguchi$^{6}$, \,Kenji Watanabe$^{6}$, \,Philip Kim$^{2,3}$,\,
Amir Yacoby$^{2,3\dag}$, \,and Ronald L. Walsworth$^{1,2,7\dag}$}

\begin{document}
\sloppy
\maketitle

\begin{affiliations}
 \item Harvard-Smithsonian Center for Astrophysics, Cambridge, MA 02138, USA
 \item Department of Physics, Harvard University, Cambridge, MA 02138, USA
 \item John A. Paulson School of Engineering and Applied Sciences, Harvard University, Cambridge, MA 02138, USA
 \item Center for Functional Nanomaterials, Brookhaven National Laboratory, Upton, NY 11973, USA
 \item Harvard College, Harvard University, Cambridge, MA 02138, USA
 \item National Institute for Materials Science, Tsukuba, Ibaraki 305-0044, Japan
 \item Center for Brain Science, Harvard University, Cambridge, MA 02138, USA
\\
$^*$ These authors contributed equally to this work.
\end{affiliations}

\begin{abstract}
The electron-hole plasma in charge-neutral graphene is predicted to realize a quantum critical system whose transport features a universal hydrodynamic description, even at room temperature~\cite{sachdev2011,levitov2016}. This quantum critical ``Dirac fluid'' is expected to have a shear viscosity close to a minimum bound~\cite{kovtun2005,muller2009}, with an inter-particle scattering rate saturating at the Planckian time $\hbar/(k_{\mathrm{B}} T)$~\cite{sachdev2011}. While electrical transport measurements at finite carrier density are consistent with hydrodynamic electron flow in graphene~\cite{bandurin2016,kumar2017,bandurin2018,berdyugin2019}, a ``smoking gun'' of viscous behavior remains elusive. In this work, we directly image viscous Dirac fluid flow in graphene at room temperature via measurement of the associated stray magnetic field. Nanoscale magnetic imaging is performed using quantum spin magnetometers realized with nitrogen vacancy (NV) centers in diamond. Scanning single-spin and wide-field magnetometry reveals a parabolic Poiseuille profile for electron flow in a graphene channel near the charge neutrality point, establishing the viscous transport of the Dirac fluid. This measurement is in contrast to the conventional uniform flow profile imaged in an Ohmic conductor. Via combined imaging-transport measurements, we obtain viscosity and  scattering rates, and observe that these quantities are comparable to the universal values expected at quantum criticality. This finding establishes a nearly-ideal electron fluid in neutral graphene at room temperature~\cite{muller2009}. Our results pave the way to study hydrodynamic transport in quantum critical fluids relevant to strongly-correlated electrons in high-$T_c$ superconductors~\cite{zaanen2018}. This work also highlights the capability of quantum spin magnetometers to probe correlated-electronic phenomena at the nanoscale.
\end{abstract}

Understanding the electronic transport of strongly-correlated quantum matter is a challenging problem given the highly entangled nature of interacting many-body systems. For example, the conventional Fermi-liquid paradigm of non-interacting quasi-particles can break down. However, if particle-particle scattering is fast compared to momentum-relaxation such that $\tau_{\mathrm{pp}}\ll\tau_{\mathrm{mr}}$, where $\tau_{\mathrm{pp}}$ is the particle-particle scattering time and $\tau_{\mathrm{mr}}$ is the momentum relaxation time, local equilibrium can be established. As a result, a simple, universal description of collective particle motion emerges in which electrical transport resembles that of hydrodynamic fluids. Due to its universality, hydrodynamics play a role in a wide range of strongly-interacting quantum matter including cold atoms~\cite{cao2011}, quark gluon plasma~\cite{shuryak2004}, and electrons in solids~\cite{gurzhi1968}.

The electron-hole (e-h) plasma in graphene at the charge neutrality point (CNP), known as the Dirac fluid~\cite{sheehy2007,muller2009,crossno2016}, is expected to share universal features present in other quantum critical systems, even at room temperature~\cite{sachdev2011}. First, particle-particle scattering occurs on the time scale $\tau_{\mathrm{pp}}\sim\tau_{\hbar}$~\cite{sachdev2011}, where the Planckian time $\tau_{\hbar} = \hbar/(k_{\mathrm{B}} T)$ is the fastest rate for entropy generation allowed by the uncertainty principle at temperature $T$. A quantum critical electron fluid involving such a similarly maximal dissipative process is postulated to underlie the linear-$T$ resistivity observed in the strange metal phase of high-$T_c$ superconductors~\cite{zaanen2018}. Second, as a result of Planckian-bounded dissipation, the Dirac fluid in graphene is expected to have a ratio of shear viscosity $\eta$ to entropy density $s$ that saturates at a lower bound $\eta/s\geq \hbar/(4\pi k_{\mathrm{B}})$~\cite{kovtun2005,levitov2016,muller2009}. Experiments with quark gluon plasma and unitary Fermi gases obtained values of $\eta/s$ that are just a few factors above this lower bound~\cite{cao2011,luzum2008}. Likewise, the Dirac fluid is predicted to come close to this ``ideal fluid'' limit~\cite{muller2009,levitov2016}. 

Compared to the non-interacting particle case, hydrodynamic electron flow is expected to be dramatically modified
due to viscosity, e.g., leading to vortices near a constriction and a parabolic velocity profile in a channel (known as Poiseuille flow)~\cite{levitov2016,torre2015}. However, while viscous flow in classical fluids and atomic gases can be directly imaged, probing hydrodynamics in electron fluids has thus far relied on indirect electrical transport measurements, requiring model-dependent interpretation. Examples include the negative vicinity voltage and channel-width-dependent resistivity in PdCoO$_2$~\cite{moll2016}, WP$_2$~\cite{gooth2018}, and graphene at finite carrier density~\cite{bandurin2016,kumar2017,bandurin2018,berdyugin2019}. As the interpretation of such indirect measurements is model-dependent, a direct, local observation (``smoking gun'') of hydrodynamic electron fluid flow is highly desirable~\cite{lucas2018}. Furthermore, while experimental evidence for the Dirac fluid in graphene was seen in violation of the Wiedemann-Franz law for a temperature range of 50 K$\lesssim T\lesssim$ 100 K~\cite{crossno2016}, transport measurements have so far been unable to provide information on whether viscous Dirac fluid flow occurs near the CNP~\cite{bandurin2016,kumar2017,bandurin2018,berdyugin2019}. Lastly, inter-species scattering (e-h) is momentum-conserving but relaxes current; whereas intra-species scattering (e-e or h-h) conserves both. Thus the question of whether there is viscous Dirac fluid flow near the CNP remains open.

Here we address this question by directly imaging the viscous Poiseuille flow of the Dirac fluid at room temperature in a graphene channel via measurement of the flow-induced stray magnetic field $\textbf{B}$ using nitrogen vacancy (NV) centers in diamond~\cite{maze2008}. Inside a conductor where the dominant scattering is momentum relaxing, for example due to impurities, phonons, or Umklapp processes, the conventional Ohmic transport of electrons in a channel exhibits a uniform current profile. In contrast, a hydrodynamic electron fluid can develop a parabolic (Poiseuille) current profile (Fig. 1a). For a 2D current distribution $\textbf{J}$, the inverse problem between $\textbf{J}$ and $\textbf{B}$ is unique~\cite{roth1989}; therefore, measurement of $\textbf{B}(x,y,z=d)$, where $d$ is the sensor-source stand-off distance, reveals the local current flow (Fig. 1b). For our experiments, magnetic imaging of current flow is performed with two complementary modalities: 1) scanning probe microscopy using a single NV center~\cite{maletinsky2012} (Fig. 1c) and 2) wide-field imaging with an ensemble of NVs~\cite{pham2011} (Fig. 1d).


We begin with scanning NV magnetic microscopy. A probe consists of a diamond nanopillar~\cite{zhou2017,xie2018} containing a single near-surface NV spin, with spatial resolution nominally set by the NV-sample distance $~\lesssim$ 50 nm. An NV spin is sensitive to the component of the magnetic field along the crystallographic axis defined by the nitrogen and the vacancy. The Biot-Savart law together with current conservation implies that the spatial-distribution of the magnetic field projection along the NV axis, $B_{||}$, contains all the information of the current distribution producing the magnetic field. For the devices we interrogate, the dimension along which the electrical current flows (the $y$-direction) is much longer than the device's narrow dimension (along the $x$-direction). Therefore, we perform a 1D scan of the single NV probe across the width of the device $W$ in order to measure $B_{||}(x)$, from which we obtain $J_y(x)$ and subsequently elucidate the presence of hydrodynamic flow.  

As a benchmark, we first perform measurements on a thin palladium channel sourced with $1\,\mathrm{mA}$ current. At each spatial location along the narrow dimension $x_n$, we perform NV optically detected magnetic resonance (ODMR) to measure the stray magnetic field $B_{||,n}$ generated by current flow $J_y$ along the $y$-direction in the palladium channel. The measured projective field $B_{||,n}$ is shown in Fig. 2a. $J_y$ is obtained by minimizing the cost function \newline
$\chi^2\equiv\sum_n|B_{||}([J_y],x_n)-B_{||,n}|^2$. Here, $\{x_n,B_{||,n}\}$ is the experimental data and $B_{||}([J_y],x)$ is the Biot-Savart functional that gives the projected field at position $x$, given a current distribution $J_y$. Details of the procedure can be found in the supplementary materials~\cite{ku2019supp}. The resulting current profile $J_y(x)$ for the Pd channel is shown in Fig. 2b. The current profile shows a clear hallmark of Ohmic transport, with a near-uniform $J_y$ that drops sharply at the edges $x/W=\pm 0.5$. 

We next apply scanning NV magnetic microscopy to measure current profiles in graphene devices, with nanoscale resolution. A device consists of an hBN-encapsulated monolayer graphene on a standard SiO$_2$/doped Si substrate. In order to stay within the linear response regime at $T=$300 K, we apply a voltage drop $\lesssim k_{\mathrm{B}}T=26\,\mathrm{mV}$ across the graphene channel, which generates a typical current $\lesssim 2\,\mu$A near the CNP. To measure the corresponding stray magnetic field $B_{||}$ with high signal-to-noise, we modulate the current at opposite polarities and perform NV spin-echo AC magnetometry~\cite{maze2008,ku2019supp}. The measured $B_{||}$ at the CNP is shown in Fig. 2a, and the resulting current density is shown in Fig. 2b. In contrast to the Ohmic profile seen in the Pd channel, the current profile in the graphene device is distinctly parabolic. Similar profiles are observed in four devices (Fig. 2c), including a measurement at large current (20 $\mu\mathrm{A}$) performed with ODMR (second profile from the left). These results constitute an observation of viscous Poiseuille flow of the Dirac fluid in graphene. Note that the profile of non-interacting electrons also develops curvature when the momentum-relaxation mean free path $l_{\mathrm{mr}}=v_{\mathrm{F}}\tau_{\mathrm{mr}}$ is comparable to the device width $W$, where the Fermi velocity $v_{\mathrm{F}}=10^6$ m/s. However, the maximum such curvature (Fig. 2b) cannot explain the experimental profile (see Supplementary Materials~\cite{ku2019supp}). Also note that we consistently observe an isolated dip in the current profile, close to the center of the graphene channel. The dip is attenuated in one graphene device (the last profile on the right of Fig. 2c), which has higher carrier density inhomogeneity as indicated by a significantly broader density-dependent resistivity (see Supplementary Materials~\cite{ku2019supp}). The nature of this current profile dip requires future investigation.



The measurement described so far assumes uniformity of the system along the $y$-direction as well as current conservation. As a complementary measurement, we perform wide-field 2D imaging of the vector magnetic field resulting from current in a graphene device, which allows us to map the electronic flow with minimal assumptions. The measurement is performed on an hBN-encapsulated graphene device with graphite top gate (optical image in the inset of Fig. 3a) fabricated on a diamond with a high-density ensemble of near-surface NV spins (Fig. 1d). We source a 100 $\mu$A current from the top contact to the bottom-left side-contact. An NV spin ensemble contains four orientations aligned with the diamond chip's four crystal axes.  Thus, vector components of the magnetic field can be reconstructed from ODMR measurements of the four NV orientations, for each pixel in the 2D image of NV fluorescence. Fig. 3a shows the measured $B_x$ and $B_y$ as a function of position $(x,y)$ in the graphene device. The current density is reconstructed from $B_x$ and $B_y$ via inverting the Biot-Savart law in Fourier space~\cite{roth1989}, resulting in the 2D vector current map shown in Fig. 3b. The measured current density has a flow pattern with direction and magnitude that are consistent with how the current is sourced and drained. More quantitatively, this vector current map allows us to verify that current conservation $\nabla\cdot\textbf{J}=0$ holds (see Supplementary Materials~\cite{ku2019supp}). Furthermore, the measured net incoming current flux at the top of the device, and outgoing flux at the bottom-left side-probe, are both consistent with a source-drain current of 100 $\mu$A.

In Fig. 3c, we show a line-cut of the current density far away from the device drain, $J_y(x,y=2.5\,\mu\mathrm{m})$. We also experimentally determine the 2D image resolution to be 420 nm, consistent with the diffraction limit of the setup (see Supplementary Material)~\cite{ku2019supp}. We can then compare the measured current density profile with calculated diffraction limited profiles of uniform and viscous electron flow, for the same total current 100 $\mu$A. As seen in Fig. 3c, the experimentally measured $J_y$ profile at the CNP clearly deviates from uniform flow and matches well a parabolic Poiseuille profile. This result provides further confirmation of the observation of viscous Dirac fluid flow in room temperature graphene.  


We next investigate the carrier density dependence of current profiles in graphene. At the CNP, an hBN-encapsulated graphene device on a SiO$_2$ substrate is stable under the optical illumination necessary for NV magnetometry, whereas away from the CNP, exposure to green light scattered from the diamond probe leads to photo-induced doping in such devices~\cite{ju2014,ku2019supp}. As a result, at any finite gate voltage, the system will equilibrate towards the CNP over time. Nevertheless, the rate is slow enough that we can obtain scanning NV magnetic measurements as a function of carrier density range~\cite{ku2019supp}. The top panel of Fig. 4a shows current profiles measured with the scanning NV magnetic microscope averaged over several different ranges of carrier densities. In all cases, the current profile is consistent with viscous flow. Similar observations are made with wide-field vector magnetic imaging. Since the hBN layer that serves as the gate-dielectric directly contacts both the graphene and the top gate, the graphene-on-diamond device is immune from photo-induced doping and allows NV measurements at a specified value of the carrier density~\cite{cadore2016}. The bottom panel of Fig. 4a shows diffraction-limited current profiles at several carrier densities up to $1.5\times 10^{12}\,\mathrm{cm}^{-2}$, all of which which are consistent with a Poiseuille profile and hence viscous flow. 

To determine the kinematic viscosity $\nu$ of the Dirac fluid, we consider an electronic flow along the $y$-direction in a 2D channel with $|x|\leq W/2$, where $W$ is the channel width. Evaluating the electronic Navier-Stokes equation~\cite{levitov2016}, assuming a no-slip boundary condition $J_y(x=\pm W/2)=0$, gives the following current density and conductivity
\begin{eqnarray}
\label{eq:profile}
J_y(x)&=&\frac{e^2v_{\mathrm{F}}\tau_{\mathrm{mr}}}{\hbar}\sqrt{\frac{n}{\pi}}E\left(1-\frac{\cosh(x/D_{\nu})}{\cosh(W/(2D_{\nu}))}\right)\,,\\
\label{eq:sigma} \sigma&=&\frac{e^2 v_{\mathrm{F}}\tau_{\mathrm{mr}}}{\hbar}\sqrt{\frac{n}{\pi}}\left(1-\frac{2D_\nu}{W}\tanh\left(\frac{W}{2D_\nu}\right)\right)\,.
\end{eqnarray}
Here, $E$ is the bias electrical field, and the Gurzhi length $D_{\nu}\equiv\sqrt{\nu\tau_{\mathrm{mr}}}$ is a characteristic length that determines the scale of viscous effects. In the limit $D_{\nu}\gg W$, $J_y$ becomes the parabolic profile of ideal viscous flow, $J_y(x)=\frac{e^2v_{\mathrm{F}}}{2\hbar\nu}\sqrt{\frac{n}{\pi}}E\left(\left(\frac{W}{2}\right)^2-x^2\right)$. As profiles with $D_{\nu}/W>0.3$ are indistinguishable from an ideal viscous flow within experimental uncertainty, we estimate the lower bound on the Gurzhi length to be $0.3W$ for the profiles presented in this paper.  Accordingly, we then obtain bounds on $\nu$ using Eq.~\ref{eq:sigma} and the measured conductivity.  Fig. 4b shows values for $\nu$ as a function of carrier density corresponding to $D_{\nu}/W=\infty$ and $0.3$, which set the upper and lower bounds respectively; as well as for $D_\nu/W =0.5$, which provides the median value. In the region near the CNP, where thermal energy becomes comparable to or exceeds the Fermi energy $k_{\mathrm{B}}T>E_{\mathrm{F}}=\hbar v_F\sqrt{\pi n}$, $\nu$ is extracted using a carrier density $n$ obtained from $E_{\mathrm{F}}(n)=k_{\mathrm{B}}T$. Away from the CNP, $\nu$ falls to between 0.1 and 0.2 m$^2$/s, consistent with the value $\approx 0.1\,\mathrm{m}^2/\mathrm{s}$ obtained close to room temperature from Refs.~\cite{bandurin2016,kumar2017}. We define the viscous scattering time as $\tau_{\nu}\equiv\nu/v^2_{\mathrm{F}}$ and show the values of $\tau_{\nu}$ normalized by the Planckian time $\tau_{\hbar}$ in the right vertical axis of Fig. 4b. $\tau_{\nu}$ is not necessarily the same as the particle-particle scattering time $\tau_{\mathrm{pp}}$, but is expected to be the same order of magnitude. The results for the viscous scattering time are consistent with the expectation for a quantum critical system, where the Plankian time sets the scale for scattering. Furthermore, the bound on $\tau_\nu$ from our experiment is consistent with the quantum critical scattering time $5\tau_{\hbar}$ measured in a Dirac fluid with terahertz spectroscopy~\cite{Gallagher2019}.


Lastly, Fig. 4c displays the shear viscosity $\eta=nm\nu$ determined from our measurements, where $m=\hbar\sqrt{\pi n}/v_{\mathrm{F}}$ is the carrier effective mass. The shear viscosity is normalized by the entropy density $s_0$ of neutral graphene at $T=300\,\mathrm{K}$, as calculated in Ref.~\cite{muller2009}. For an ``ideal fluid", this quantity reaches the conjectured lower bound $\eta/s=\hbar/(4\pi k_{\mathrm{B}})=0.08\,\hbar/k_{\mathrm{B}}$. Close to the CNP, we observe $\eta/s_0\approx 0.3-0.8\,\hbar/k_{\mathrm{B}}$, which lies between the theoretical value $0.26\,\hbar/k_{\mathrm{B}}$ for a Dirac fluid at 300 K from Ref.~\cite{muller2009} and an experimental estimate $\approx 10\,\hbar/k_{\mathrm{B}}$ for a Dirac fluid at $\sim$100 K from Ref.~\cite{lucas2016}. We note that our value for $\eta/s_0$ near the CNP is comparable to the value $\approx 0.4\,\hbar/k_{\mathrm{B}}$ obtained in cold atoms~\cite{cao2011} and $\lesssim 0.5\,\hbar/k_{\mathrm{B}}$ in quark gluon plasma~\cite{luzum2008}. Thus, our result adds a new data point for $\eta/s$ in strongly-interacting quantum matter, the first from a condensed matter system, and highlights the trend of strongly-interacting systems exhibiting $\eta/s$ within an order of magnitude of the ``ideal fluid'' lower bound.

In conclusion, we directly imaged the viscous flow of the Dirac fluid in graphene at room temperature using a quantum spin magnetometer, and established the Dirac fluid as a quantum-critical, nearly-ideal electron fluid. While previous indirect experiments in graphene employing electrical transport implied the strongest hydrodynamic effect at intermediate temperatures $\sim$100-200 K~\cite{bandurin2016,kumar2017}, we find that viscous flow is robust at room temperature, both near and far from the charge-neutrality point. This result has implications for the characterization of graphene devices at elevated temperatures: the conventional procedure using the Drude conductivity cannot be justified, as pointed out in Ref.~\cite{bandurin2016} and emphasized here. Building on our work, it will be interesting to image viscous flow in graphene at different temperatures to extract the viscosity $\nu(n,T)$, which will serve as a benchmark for testing many-body theories. Electronic turbulence may also be explored via magnetic imaging experiments, as one can tune the Reynolds number by increasing the current and observe the evolution of the time-averaged flow profile from laminar to turbulent. Lastly, our magnetic imaging techniques can be deployed for nanoscale study of an increasing number of materials manifesting viscous electronic flow, as well as other correlated electronic phenomena such as topological transport in the quantum spin Hall effect~\cite{nowack2013}.
 
\clearpage 
 
\includegraphics[trim=96 20 50 130,clip,scale=1]{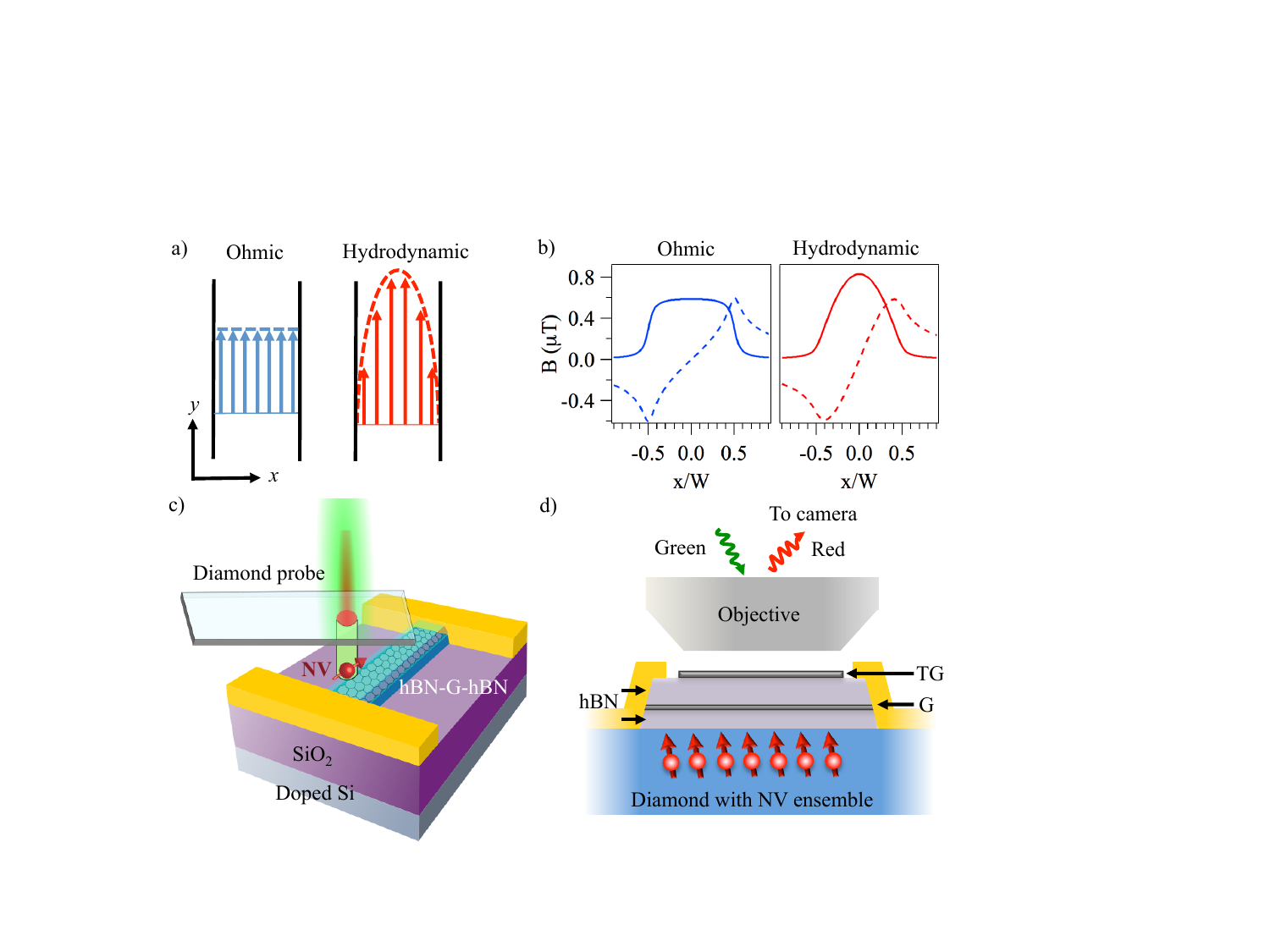}\\
\noindent {\bf Fig. 1. Probing viscous electronic transport via magnetic field imaging.}   ({\bf a}) Local current density $J_y$ in a 2D conductor oriented along $y$-direction for conventional Ohmic transport (left) and viscous Poiseuille flow (right). ({\bf b}) Magnetic field $B_x(x)$ (solid line) and $B_z(x)$ (dashed line) at $z=d$ generated by an electronic flow $J_y(x)$ in the $y$-direction for Ohmic (left) and viscous Poiseuille flow (right). Here, the magnetic field profile is shown for a channel width $W= 1\,\mu\mathrm{m}$, current $I= 1\,\mu\mathrm{A}$, and stand-off distance $d=50\,\mathrm{nm}$. Hence, a quantum spin magnetometer located at $z = d$ from a 2D current distribution $\textbf{J}$ probes the local current via sensing the local magnetic field $\textbf{B}$ generated by $\textbf{J}$. ({\bf c}, {\bf d}) Magnetic field imaging modalities employed in this work. ({\bf c}) Scanning NV magnetometry with a diamond probe containing a single near-surface NV spin. NV red photoluminence (PL) is collected by a confocal microscope. hBN-encapsulated graphene device (hBN-G-hBN)~\cite{dean2010} is fabricated on a SiO$_2$/doped Si substrate. Typical stand-off distance $\lesssim 50\,\mathrm{nm}$ between NV and graphene. ({\bf d}) Wide-field magnetic imaging. An hBN-encapsulated graphene device with graphite top gate (TG) is fabricated on a macroscopic diamond chip containing a 	high-density ensemble of near-surface NVs. Green excitation light is sent through an objective to illuminate the field of view. NV PL is collected by the same objective and imaged onto a camera with about $400\,\mathrm{nm}$ 2D resolution.

\clearpage
\noindent{
\begin{centering}
\includegraphics[trim=20 0 0 0, clip, scale=1]{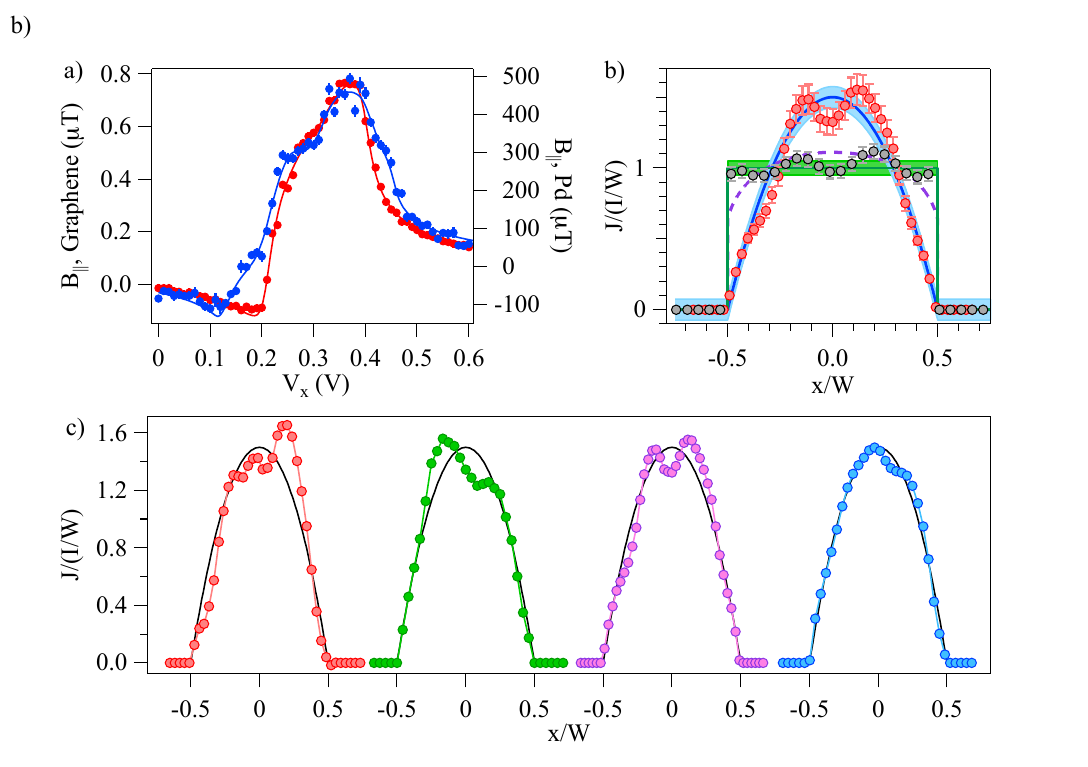}
\end{centering}}\\
\noindent {\bf Fig. 2. 1D scanning NV microscopy of Ohmic flow in a metallic channel and viscous flow in graphene at the CNP.} ({\bf a}) Scanning magnetometry of the projective stray field $B_{||}$ measured with the Pd channel (red points, units shown in the right vertical axis) and the graphene device at the CNP point (blue points, units shown in the left vertical axis). Horizontal axis is the piezo voltage that drives the scanning probe in the $x$-direction. Lines: $B_{||}$ from reconstructed current density $J_y$ that minimizes the cost function $\chi^2$. Data for the Pd channel is measured with optically-detected magnetic resonance (ODMR) and a sourced current of 1 mA. Data for graphene is measured with spin-echo AC magnetometry and a source-drain voltage $V_{\mathrm{sd}}$=5.8 mV. Insets: optical image of graphene device (left) and atomic force microscopic image of Pd channel (right); the width is 1 $\mu\mathrm{m}$ and $800\,\mathrm{nm}$, respectively. Pd channel is 30 nm thick and $100\,\mu\mathrm{m}$ long. The graphene device whose data is shown in (a) and (b) is $6\,\mu\mathrm{m}$ long, whereas the graphene devices measured with scanning magnetometry (shown in (c)) have length $\geq 5\,\mu\mathrm{m}$. For all devices, scanning magnetometry is performed at a $y$ position near the mid-point along the length of the channel. For graphene devices, we ensure the $y$ position of the scan is within $1\,\mu\mathrm{m}$ of the longitudinal mid-point via atomic force microscopy with the diamond probe.  ({\bf b}) Reconstructed current density $J_y(x)$. $J_y$ is normalized by the average charge carrier flux $I/W$, where $I=\int\mathrm{d}x\,J_y(x)$ is the total flux and $W$ is the width of the channel. The spatial coordinate $x$ is normalized by $W$ and centered on the channel.  Red points: graphene at the CNP. Gray points: Pd channel. Error bars correspond to the relative deviation of $J_y$ that generates $2\chi^2$, where $\chi^2$ is the cost function. Blue (green) lines: ideal viscous (uniform) flow with 5\% error band. Purple dashed curve: current profile of non-interacting electrons with diffusive boundary condition and momentum-relaxing mean free path $l_{\mathrm{mr}}=0.625W$. ({\bf c}) Current profile at the CNP from four different graphene devices. Second profile from left is for current of 20 $\mu\mathrm{A}$ and is measured with ODMR, whereas other profiles have current $\lesssim 2\,\mu$A and are measured with spin-echo AC magnetometry

\clearpage

\noindent{
\begin{centering}
\includegraphics[trim=130 90 120 110, clip, scale=1]{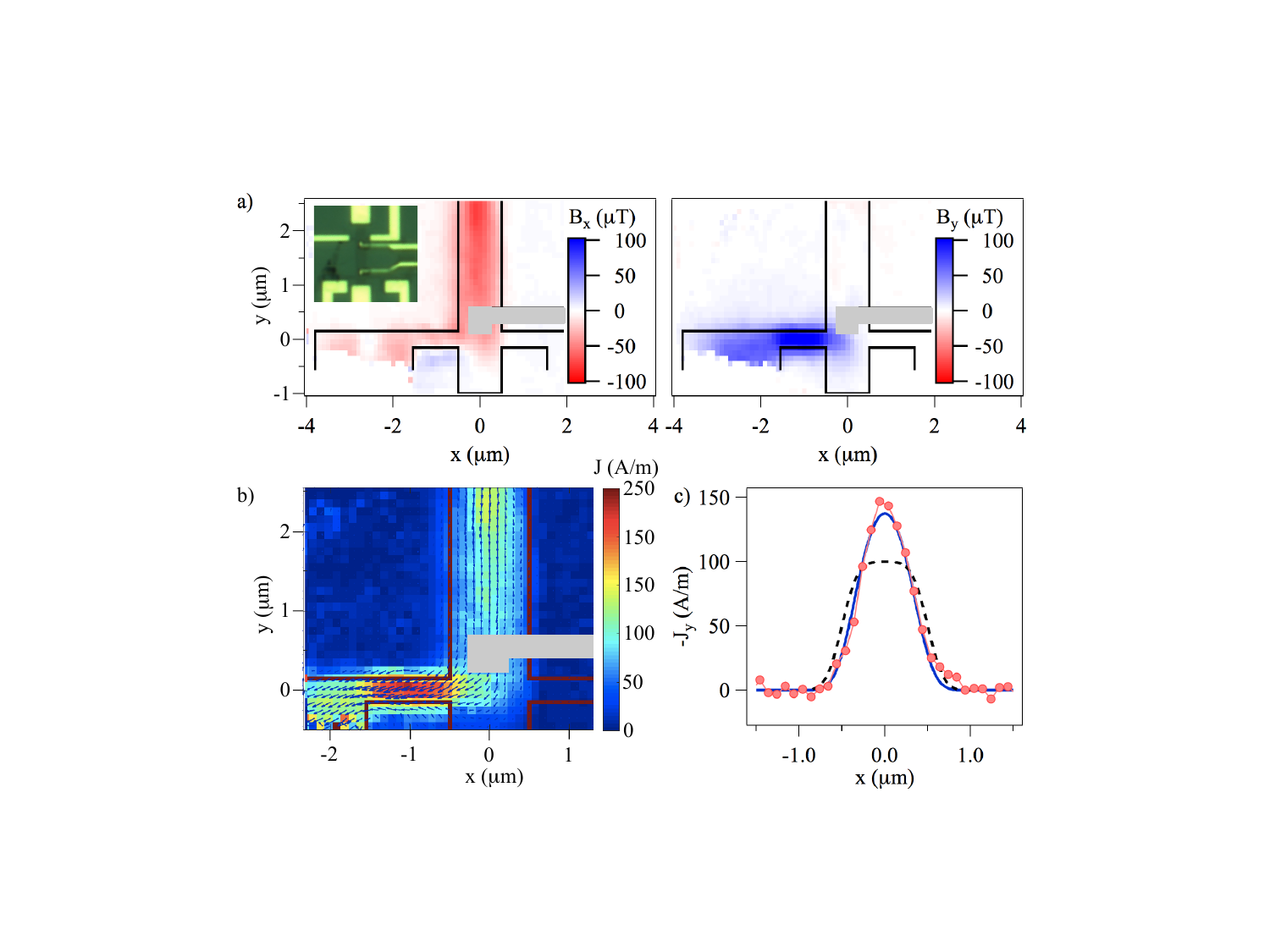}
\end{centering}}\\
\noindent {\bf Fig. 3. Wide-field imaging of viscous electron flow in graphene.} ({\bf a}) 2D spatial distribution of $B_x$ (left) and $B_y$ (right) measured by wide-field NV vector magnetometry, for source-drain current of 100 $\mu\mathrm{A}$ and NV ensemble imager located in positive $z$-direction. Inset: optical image of device channel $W=1\,\mu\mathrm{m}$.  Current is sourced from top contact to bottom-left contact. ({\bf b}) Vector (arrow) and amplitude (color) plots of current density $\textbf{J}(x,y)$ extracted from vector field measurement. In ({\bf a}) and ({\bf b}), gray outline denotes area covered by a metallic top-gate contact that obstructs light. ({\bf c}) Horizontal line-cut of ({\bf b}) far away from drain, $J_y(x, y=2.5\,\mu\mathrm{m})$, measured at the CNP. Shown also are calculated diffraction-limited profiles of a uniform (black) and viscous (blue) flow. 

\clearpage
\noindent{
\begin{centering}
\includegraphics[trim=180 0 0 0, clip, scale=1]{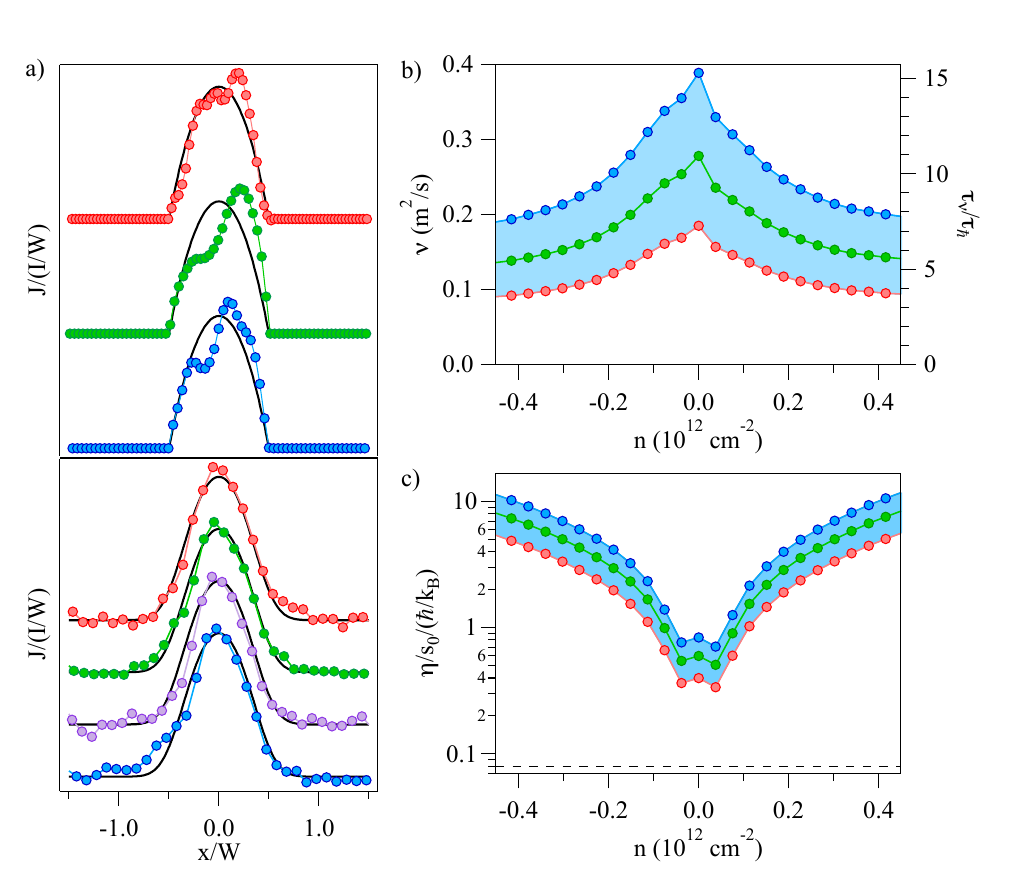}
\end{centering}}\\
\noindent {\bf Fig. 4. Electron hydrodynamics in graphene near charge neutrality point (CNP).} ({\bf a}) Normalized current density $J_y/(I/W)$ as a function of $x/W$, where $I$ is the total charge carrier flux and $W$ is the graphene device width. Upper panel: data obtained from scanning NV probe magnetometry. Data points are for the CNP (top), averaged over carrier density $n=0.1-0.3\times10^{12}\,\mathrm{cm}^{-2}$ (middle), and averaged over $n=0.3-1.5\times10^{12}\,\mathrm{cm}^{-2}$ (bottom). Solid curve is for calculated ideal viscous flow. Bottom panel: data obtained from wide-field NV vector magnetic imaging. Data points from top to bottom correspond to the CNP and $n=0.17\,,0.85\,,1.5\times 10^{12}\,\mathrm{cm}^{-2}$, respectively. Solid curve is for calculated diffraction-limited ideal viscous flow. ({\bf b}) Bounds on kinematic viscosity $\nu$ obtained from conductivity. $D_{\nu}/W=\infty$, 0.3, and 0.5 set the upper bound, lower bound, and median, respectively. Shown on right vertical axis is the corresponding viscous scattering time $\tau_{\nu}$ normalized by the Planckian time $\tau_{\hbar}$. ({\bf c}) Ratio of shear viscosity $\eta$ to $s_0$, where $s_0$ is entropy at the CNP calculated in~\cite{muller2009}. $\eta/s_0$ is normalized by $\hbar/k_{\mathrm{B}}$. Black dashed line is the ``ideal fluid'' lower bound~\cite{kovtun2005} $1/(4\pi)$.

\clearpage

\begin{methods}



\subsection{NV spin physics.}
An NV spin consists of a substitutional nitrogen atom and an adjacent vacancy in the diamond lattice. Its ground state is an $S=1$ electronic spin-triplet with $m_s=0,\,\pm 1$ eigenstates, each with magnetic moment $m_s\gamma_e$, where $\gamma_e=$2.8 MHz/G. An NV spin can be initialized into the $m_s=0$ state with optical excitation at around 532 nm wavelength. Optical excitation also generates spin-dependent photoluminence (PL) in the 640-800 nm wavelength range, which allows optical spin readout. The $m_s=\pm 1$ states are split from the $m_s=0$ state by about 2.87 GHz at zero magnetic field, and are further split from each other at finite magnetic field due to the Zeeman effect. 

Experiments employ electronic grade single crystal diamonds with  \{110\} cut from Element Six, created using chemical vapor deposition. Atomic force microscopy ensures surface roughness is under 1 nm. Near-surface NV spins are created using $^{15}$N implantation (Innovion) at 6 keV followed by annealing for both wide-field imaging and scanning probe magnetometry. This implantation energy generates NV spins at depth $\sim$ 10-20 nm beneath the diamond surface. For wide-field imaging, a dosage [N]$=2\times10^{13}\,\mathrm{cm}^{-2}$ leads to an ensemble of [NV]$\sim10^{12}\,\mathrm{cm}^{-2}$. For scanning probe magnetometry, a dosage is selected to ensure each probe contains on average one NV spin~\cite{zhou2017}. The diamonds are cleaned using a mixture of sulfuric, nitric, and perchloric acids prior to deposition of van der Waals heterostructure or fabrication of scanning probe.

\subsection{Device Fabrication.}
A polymer-free assembly method~\cite{wang2013} is used to prepare a van der Waals heterostructure for the devices. hBN and monolayer graphene (G) flakes are exfoliated on standard SiO$_2$/p-doped Si substrate. Polymer polycaprolactone~\cite{shin2019} is used to pick up the flakes. For scanning measurement, three-layer hBN-G-hBN stacks are prepared on a standard substrate; the thickness ranges from 20-50 nm for the bottom hBN and 10-20 nm for the top hBN. For wide-field imaging, a four-layer structure, consisting of an hBN-encapsulated graphene stack (top hBN 13(1) nm and bottom hBN 27(1) nm thick) together with a graphite top gate is deposited on a diamond.

Electron-beam lithography (EBL) is performed using Elionix 7000 with 495 PMMA C6 as resist. In order to mitigate charge build-up during fabrication of the graphene-on-diamond device, an additional layer of Aquasave is spun onto the diamond and conductive carbon tape is attached to the diamond edges. To pattern the stack and to make edge contacts, STS ICP reactive ion etch (RIE) with CHF$_3$ chemistry is employed. Electron-beam evaporation deposits 10 nm of Cr as adhesion layer followed by Au of minimum 100 nm thickness for metallization and creation of other electrical structures (e.g., microwave delivery for the graphene-on-diamond device and bond pads).

For the graphene-on-diamond device, it is necessary to create contact to the top gate graphite without shorting to the graphene. To accomplish this, first EBL is used to write a dielectric structure with hydrogen silsesquioxane (HSQ), which serves to insulate the top gate contact from the graphene. The top gate contact is then created. It is also necessary to ensure the top gate graphite is not shorted to graphene edge contacts. To accomplish this, graphite in the vicinity of the edge contacts is etched away using RIE with O$_2$ chemistry, which does not affect hBN.

\subsection{Wide-field Magnetic Imaging.}
The sample is positioned on a combined translational stage of PI (Physik Instrumente) M686.D64 for coarse positioning with $\sim0.1\,\mu$m precision, and P-541.2DD for fine positioning with $\sim$nm precision. NV centers are optically excited with a 532 nm laser (Coherent COMPASS 315M-100) switched on and off by an acousto-optic modulator (Isomet 1250C-974). The light is directed onto the sample with an Olympus 0.9NA 100x objective. A Kohler-illumination system expands the laser beam to illuminate an area of about 5 $\mu$m in diameter on the sample. About 10 mW of excitation illuminates the sample. NV photoluminence (PL) is collected by the same objective. The collected light passes through a 552 nm edge dichroic (Semrock LM01-552-25), after which it is separated from the excitation, passes through another 570 nm long-pass filter to further reduce light not in the PL wavelength range, and is imaged via an f=200 nm tube lens onto a camera (Basler acA1920-155um). The output of a microwave (MW) synthesizer (SRS SG384) is controlled by a switch (Mini-Circuits, ZASWA-2-50DR+), then amplified (Mini-Circuits, ZHL-16W-43-S+), and delivered onto the PCB housing the sample. A permanent magnet, whose position is controlled by three sets of motorized translational stages (Thorlabs MTS25-Z8), is used to apply a uniform external magnetic field. A gate voltage is supplied by a Keithley 2420, and DC current for the measurement is supplied by a Keithley 6221.

ODMR spectra are measured by capturing an NV PL image at different MW frequencies. The spectra contain both the $m_s=0\Leftrightarrow-1$ (lower) and $+1$ (upper) resonances. To extract the resonance frequencies $f_{\pm}$, Lorentzian fits of PL vs frequency are performed at each pixel; the fitting function consists of a single Lorentzian or double-Lorentzian if the $^{15}$N hyperfine splitting is resolved. This procedure is applied with the external field ($\approx$ 90G) aligned along three of the four NV axes, labeled as $i=1,2,3$. At each NV orientation, resonance frequencies for the upper transition $f_+(\pm I)$ and lower transition $f_-(\pm I)$ are recorded for current  through the sample in the positive and negative polarity $\pm I$.   At a given NV orientation, the projected stray field generated by the current can then be obtained as $B_{||}=[(f_+(+I)-f_+(-I))-(f_-(+I)-f_-(-I))]/(4\gamma_e)$. Determination of three projective field components $B_{||,i}$ allow the vector magnetic field in the NV layer to be determined for all three Cartesian directions.

Four classes of orientations exist in the NV ensemble. Up to a rotation in the $xy$ plane, the orientation vectors are given by
\begin{eqnarray}
 \hat{u}_{1,\mathrm{NV}}&=&\frac{1}{\sqrt{3}}(\sqrt{2},0,1)\,,\\
 \hat{u}_{2,\mathrm{NV}}&=&\frac{1}{\sqrt{3}}(-\sqrt{2},0,1)\,,\\
 \hat{u}_{3,\mathrm{NV}}&=&\frac{1}{\sqrt{3}}(0,\sqrt{2},1)\,,\\
 \hat{u}_{4,\mathrm{NV}}&=&\frac{1}{\sqrt{3}}(0,-\sqrt{2},1)\,.
\end{eqnarray}
Here, the coordinates are defined by the diamond, which we call the NV frame, $(x_{\mathrm{NV}},y_{\mathrm{NV}},z)$. This differs from the coordinate in the device frame $(x,y,z)$, which is such that the $y$-direction is along the length of the channel, by a rotation of angle $\theta$ in the $xy$ plane. For a vector measurement, three projected fields are measured, $B_i=\textbf{B}\cdot\hat{u}_{i,\mathrm{NV}}$. Then, the vector field in the NV-frame can be determined:
\begin{eqnarray}
 B_z&=&(\sqrt{3}/2)(B_1+B_2)\,,\\
 B_{x,\mathrm{NV}}&=&\sqrt{3/2}(B_1-B_2)/2\,,\\
 B_{y,\mathrm{NV}}&=&\sqrt{3/2}(B_3-B_z/\sqrt{3})\,.
\end{eqnarray}
The vector field in the device frame can be obtained by a rotation angle $\theta$,
\begin{equation}
\left(
\begin{array}{c}
B_x\\
B_y
\end{array}
\right)
=
\left(
\begin{array}{cc}
\cos\theta&-\sin\theta\\
\sin\theta&\cos\theta\\
\end{array}
\right)
\left(
\begin{array}{c}
B_{x,\mathrm{NV}}\\
B_{y,\mathrm{NV}}
\end{array}
\right)\,.
\end{equation}

Current reconstruction is done using direct inversion of the Biot-Savart law in Fourier space,
\begin{eqnarray}
b_x(\textbf{k},d)&=&\frac{\mu_0}{2}e^{-d|\textbf{k}|}j_y(\textbf{k})\,,\\
b_y(\textbf{k},d)&=&-\frac{\mu_0}{2}e^{-d|\textbf{k}|}j_x(\textbf{k})\,.
\end{eqnarray}
where $\mu_0$ is the vacuum permeability, and the stand-off distance used is $d=$50(10) nm~\cite{ku2019supp}. In this modality, uncertainty in the stand-off distance is the dominant source of error for the current density, which is about 9 A/m.

Optical diffraction limits the point spread function of the wide-field magnetic imager, placing a Fourier low-pass filter on the current reconstruction, with a cutoff spatial frequency significantly lower than the cutoff placed by the NV layer stand-off distance (i.e., the imager spatial resolution $\sim$400 nm is large compared to the standoff $d\sim$ 50 nm). Thus, high spatial frequency noise is suppressed by the wide-field magnetic imager and direct inversion of the device current is appropriate.

\subsection{Scanning NV Magnetic Microscopy.}
The setup is described in Ref.~\cite{maletinsky2012}. Scanning of the sample is performed with piezoelectric nanopositioners Attocube ANPxyz101 and ANSxyz100. The measurement pulse sequence is controlled by a Tektronic Arbitrary Waveform Generator (AWG) 5014C. Microwaves are supplied by a Rhode Schwartz SMB100A. Since the device for scanning measurement is fabricated on a substrate with a global back gate, microwaves cannot be delivered on-chip as capacitive effects lead to poor microwave transmission. Instead, we use a separate antenna made from a wire of about 20 $\mu$m diameter placed about 50 $\mu$m above the device. The movement of the antenna is controlled by manual translational stages. The diamond probe is described in Ref.~\cite{zhou2017,xie2018}. 

A scanning measurement consists of iterating the piezo voltage $V_x$ of the sample stage, and then performing OMDR or spin-echo magnetometry  at each value of $V_x$. For a scanning ODMR measurement, current is supplied by a Keithley 2420. The ODMR measurement is performed in the same manner as in wide-field magnetic imaging, except that only one projection of $B_{||}$ is measured.  For spin-echo AC magnetometry, a Hahn spin-echo $(\pi/2)_x-\tau-(\pi)_x-\tau-\pm(\pi/2)_{x,y}$ is employed. The first $(\pi/2)_x$ pulse generates an equal NV superposition state $(|0\rangle+|-1\rangle)/\sqrt{2}$. A modulating square pulse of bias  $\pm V_{\mathrm{AC}}$ is supplied to the graphene device from the AWG during the NV free precession time, generating a field of $\pm B_{\mathrm{AC}}$ on the NV. During the free precession, the NV picks up a total phase of $\phi=\pi-2\gamma_{\mathrm{e}}\tau B_{\mathrm{AC}}$. The NV is therefore in the superposition state $(|0\rangle+e^{i\phi}|-1\rangle)/\sqrt{2}$. The last $\pm(\pi/2)_{x,y}$ pulse leads to a PL readout signal of $\mathrm{PL}_{\pm x}=\mathrm{PL}_0\pm A\cos(2\phi)$ for $\pm(\pi/2)_x$ and $\mathrm{PL}_{\pm y}=\mathrm{PL}_0\pm A\sin(2\phi)$ for $\pm(\pi/2)_y$. Here, $\mathrm{PL}_0+A$ is the PL signal for $m_s=0$ state and $\mathrm{PL}_0-A$ is the PL signal for $m_s=-1$ state. Therefore, the NV phase can be determined as
\begin{equation}
\phi=\tan^{-1}\left(\frac{\mathrm{PL}_{+y}-\mathrm{PL}_{-y}}{\mathrm{PL}_{+x}+\mathrm{PL}_{-x}}\right)\,.
\end{equation}

The microwave pulses are tuned to the $m_s=0\leftrightarrow -1$ transition as determined via ODMR. Before the start of a scan, we spin-echo sequence is run without $V_{\mathrm{AC}}$ to determine the Hahn echo spin coherent time $T_2$, as well as locations of the NV spin coherence revival in the presence of the natural abundance $^{13}$C bath. $\tau\approx 10$ or $20\,\mu\mathrm{s}$ is used. During a scan, the microwave drive Rabi frequency is measured at each spatial location to to calibrate the microwave pulses. Spin-echo measurements are then performed with all four possible final microwave pulses $\pm(\pi/2)_{x,y}$ to obtain a measurement of the NV phase as described above. Differential measurements are also performed with the AC bias sequence $+V_{\mathrm{AC}},-V_{\mathrm{AC}}$ and $-V_{\mathrm{AC}},+V_{\mathrm{AC}}$, leading to a total of 8 pulse sequence combinations. 

\subsection{Current Reconstruction from Scanning NV Microscopy.}
Scanning NV microscopy measures only one projection of the magnetic field generated by current through the device. However, the Biot-Savart law together with current conservation, which has the form $k_x j_x+k_yj_y=0$ in Fourier space, requires the map of one projection of the magnetic field to contain all information of a 2D current distribution. In our experiment, the NV has an orientation vector \\
$\hat{u}=(\sqrt{2/3}\cos\theta,\sqrt{2/3}\sin\theta,\sqrt{1/3})$, where $\theta$ is the angle between the NV orientation projected on the $xy$ plane and the $x$-axis. Here, the device being studied is used to define the coordinate axis, with the long direction of the device taken as the $y$-direction. Therefore, the projected field is $B_{||}(x)=\sqrt{2/3}\cos\theta B_x(x)+\sqrt{1/3}B_z(x)$. The sample is placed on the setup in such a way as to minimize $\theta$.

There are two challenges with applying the Fourier inversion method described previously to scanning measurements. First, $B_{||}$ involves $B_z$, which has a long $\sim1/x$ tail. Direct inversion requires measurement over a large field of view or otherwise leads to long-wavelength artifacts (e.g., offsets or a slope), which in general can be corrected but nevertheless are inconvenient.  Second, current reconstruction involves inverting a low-pass filter which, in the absence of other low-pass filter such as optical diffraction, leads to amplification of noise at high frequencies. Therefore, regularization is required~\cite{meltzer2017}. However, regularization needs to be performed with care: a naive application of a global regularization can smoothen out a sharp feature, such as at the edges.

The goal is to have a current reconstruction procedure that is generally applicable to an arbitrary current profile $J_y(x)$ flowing in a channel oriented along the $y$-direction. First, the current density is parameterized as $J_y=J_y(\{a_n\},x)$, with $\{a_n\}$ being free parameters that determine $J_y$. It is convenient to capture the expected overall shape of the current profile with a minimal number of parameters in a function called $J_0$. For this purpose, the functional form of Eq.~\ref{eq:profile} is chosen, which can vary from a parabola to a rectangular function. While $J_0$ is the solution of the electronic Navier Stoke equation, it is used here merely to set the functional form without assuming hydrodynamics. The second part of the parameterization $\Delta J$ captures the remaining variation; it is a linear interpolation with $N$ equally spaced points inside the channel and parametrized by the values $a_n=\Delta J\left(x_n\right)$, where $x_n\equiv-\frac{W}{2}+n\frac{W}{N}$ for $n=0,\,1,\,...,\,N$ is the $n^{\mathrm{th}}$ equally-spaced point along the width of the channel. Since the behavior at the edges is already captured by $J_0$, it is allowed to set $a_0,a_N=0$. The mathematical expression for the parameterization is
\begin{eqnarray}
J_y(x,W,D,\{a_n\})&=&A_0J_0(x,W,D)+\Delta J(x,W,N,\{a_n\})\,,\\
J_0(x,W,D)&\equiv&\left(1-\frac{\cosh(x/D)}{\cosh(W/(2D))}\right)\Pi(x/W)\,,\\
\Delta J(x,W,N,\{a_n\})&\equiv&\sum^{N}_{n=1} J_{\mathrm{lin}}(x-(x_{n-1}+x_n)/2,W/N,a_{n-1},a_n)\,,\\
J_{\mathrm{lin}}(x,\Delta W,a,b)&\equiv& \left(
\left(\frac{ b-a}{\Delta W}\right)x+\frac{a+b}{2}
 \right)\Pi(x/\Delta W)\,.
\end{eqnarray}
Here, $\Pi(x)$ is the rectangular function, which is unity for $|x|<1/2$ and zero otherwise. Note that no particular model is assumed for the current density; also the parameterization described above allows $J_y$ to take on any functional form. The only prior knowledge exploited in this procedure is the fact that the current density is zero outside the channel.

With this parameterization, functional can be constructed that generates $B_{||}$ for a given $J_y$:
\begin{eqnarray}
\nonumber B_{||}(x,x_{\mathrm{ctr}},W,D,\{a_n\},d,\theta,B_0)&=&B_0+\sqrt{\frac{2}{3}}\cos(\theta) \alpha_x(x,d)\ast J_y(x-x_{\mathrm{ctr}},W,D,\{a_n\})\\
&+&\frac{1}{\sqrt{3}}\alpha_z(x,d)\ast J_y(x-x_{\mathrm{ctr}},W,D,\{a_n\})\,.
\end{eqnarray}
Here, $x_{\mathrm{ctr}}$ is the center of the channel, $d$ is the stand-off distance, $B_0$ is an offset that accounts for potential contribution in $B_z$ from far away current flowing not strictly in the $y$-direction, $\ast$ denotes a 1D convolution, and the kernels are 
\begin{eqnarray}
\alpha_x(x,d)&\equiv&\frac{\mu_0}{2\pi}\frac{x}{(d^2+x^2)}\,\\
\alpha_z(x,d)&\equiv&-\frac{\mu_0}{2\pi}\frac{d}{(d^2+x^2)}\,.
\end{eqnarray}
To obtain the parameters $\{x_{\mathrm{ctr}},W,D,\{a_n\},d,\theta,B_0\}$, the cost function is minimized:
\begin{equation}
\chi^2=\sum_i |B_{||}(x_i,x_{\mathrm{ctr}},W,D,\{a_n\},d,\theta,B_0) -B_{||,i} |^2\,,
\end{equation}
where $\{x_i,B_{||,i}\}$ is the experimental data. 

Inverting the Biot-Savart law, which acts as a low-pass filter, without regularization leads to unphysical amplification of high-frequency noises on the magnetic field, and therefore regularization is routinely required for the inverse problem. Here, regularization is performed by subjecting the $\chi^2$ minimization to a global constraint $|\Delta J''|<\Delta J_{\mathrm{max}}''$, where $\Delta J''$ is the (numerical) second derivative of $\Delta J$. This is equivalent to a low-pass filter of the form $k^{-2}$ in Fourier space. The constraint $\Delta J_{\mathrm{max}}''$ is chosen such that the reduced $\chi^2$ is close to unity, $\chi^2_\nu\equiv\chi^2/[\delta B^2(N-N_{\mathrm{p}})]\approx 1$. Here, $\delta B$ is the typical error bar of the data, $N$ is the number of data points and $N_{\mathrm{p}}$ are the number of free parameters in the functional.  We note that $J_0$ itself is well-behaved, and therefore only regularization on $\Delta J$ is necessary.

\subsection{Transport Measurement.}
Transport measurement of device conductivity $\sigma$ is performed with an SRS lock-in amplifier (SRS SG384) at 17.77 Hz using a current-bias modality, with a large attached load $ \approx$10 M$\Omega$ and source $\approx$100 nA. The gate voltage $V_g$ is supplied by a Keithley 2420. Carrier density $n$ is determined using $n=C_g(V_g-V_D)$, where $V_D$ is the location of the CNP determined from location of the peak resistivity $\rho$, and $C_g=\varepsilon_0\varepsilon_r/(et)$ is the gate capacitance per area, with $\varepsilon_0$ the permittivity of free space, $\varepsilon_r$ the dielectric constant, and $t$ the thickness of the dielectric. For the graphene-on-diamond device, $\varepsilon_r=4$ and $t=\,13\mathrm{nm}$ are used for the hBN gate dielectric. For the devices made on the standard SiO$_2$/Si substrates, $\varepsilon_r=3.9$ and $t=385\,\mathrm{nm}$ are used.

\end{methods}

\clearpage



\begin{addendum}
 \item This material is based upon work supported by, or in part by, the United States Army Research Laboratory and the United States Army Research Office under Contract/Grants No. W911NF1510548 and No. W911NF1110400. This work is supported by the U.S. Department of Energy, Office of Science, Office of Basic Energy Sciences Energy Frontier Research Centers program under Award Number DE-SC-0001819. It is also supported in part by the Army Research Office under Grant Number W911NF-17-1-0023 and W911NF-17-1-0574; The STC Center for Integrated Quantum Materials, NSF Grant No. DMR-1231319; The NSF under Grant No. EFMA-1542807; and the Elemental Strategy Initiative conducted by MEXT, Japan, and JSPS KAKENHI grant JP15K21722 (K.W. and T.T.). Finally this work was partly carried out at the Aspen Center for Physics, which is supported by National Science Foundation grant PHY-1607611. F.C. acknowledges support from the Swiss National Science Foundation grant no. P300P2-158417. This research used resources of the Center for Functional Nanomaterials, which is a U.S. DOE Office of Science Facility, at Brookhaven National Laboratory under Contract No. DE-SC0012704. This work was performed, in part, at the Center for Nanoscale Systems (CNS), a member of the National Nanotechnology Infrastructure Network, which is supported by the NSF under award no. ECS-0335765. CNS is part of Harvard University.
 \item[Competing Interests] The authors declare that they have no
competing financial interests.
 \item[$^\dag$Correspondence] Correspondence and requests for materials
should be addressed to Amir Yacoby \\(yacoby@g.harvard.edu) and Ronald L. Walsworth (rwalsworth@cfa.harvard.edu).
\end{addendum}


\end{document}